\documentclass[twocolumn,prl,showpacs,preprintnumbers,amsmath,amssymb]{revtex4-1}
\usepackage{dcolumn}
\usepackage{graphicx}
\usepackage{color}

\begin{document}

\author{Shaon Sahoo}
\affiliation{Department of Physics,
 Indian Institute of Science, Bangalore 560012, India}

\title{Entanglement Localization and Optimal Measurement}
\begin{abstract}
The entanglement can be localized between two noncomplementary parts of a
many-body system by performing measurements on the rest of the system.
This localized entanglement (LE) depends on the chosen basis set of measurement
(BSM). We derive here a generic optimality condition for the LE, which, besides
being helpful in studying tripartite systems in pure states, can also be of use
in studying mixed states of general bipartite systems. We further discuss a
{\it canonical} way of localizing entanglement, where the BSM is not chosen
arbitrarily, but is fully determined by the properties of the system. The LE
obtained in this way, we call the {\it localized entanglement by canonical
measurement} (LECM), is not only operationally meaningful and easy to calculate
in practice (without needing any demanding optimization procedure), it provides
a nice way to define the {\it entanglement length} in many-body systems.
For spin-1/2 systems, the LECM is shown to be optimal in some important cases.
At the end, some numerical results are presented for $j_1-j_2$ spin model to
demonstrate how the LECM behaves.
\end{abstract}

\maketitle

Besides its importance in interpreting and understanding quantum mechanics,
the entanglement has gained immense interest in recent times as it has the
potential to play a significant role in modern technology. In addition, it
has become an important tool to study quantum many-body systems \cite{amico}.
Some of the very useful measures used here for studying quantum non-local
nature of a system are pairwise entanglement \cite{venuti,osterloh}, local
entropy \cite{sjgu}, localizable entanglement \cite{popp} and negativity
\cite{vidal}. Here we study entanglement localization which is important for
two reasons -from a practical point of view, it can be a useful method of
producing entangled pairs (especially from three-body systems) and secondly,
we get an alternative theoretical way for studying quantum many-body systems.
Here it may be stressed that, almost all the measures which try to quantify
mutual quantum behavior between two disjoint parts of a many-body
system are either
difficult to calculate (often needing optimization procedure) or they do not
have any operational meaning. The measure we present here (LECM) is not only
easy to calculate, it also has some operational meaning.

\begin{figure}[t]
\includegraphics[width=8.0cm]{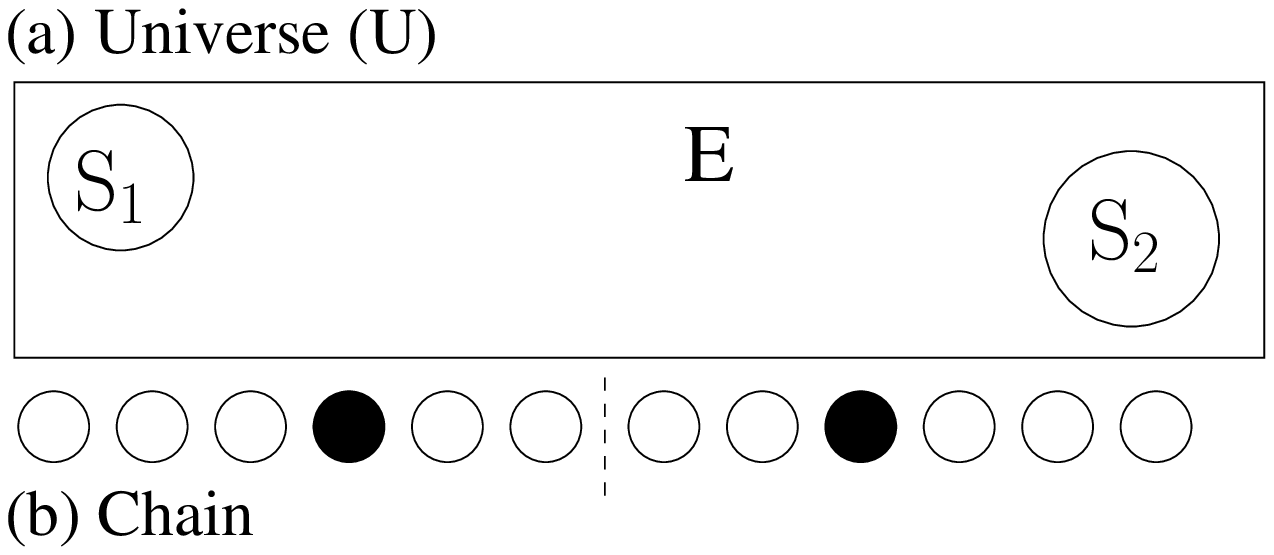}
\caption{\small (a) Total system or Universe ($U$): $S_1$ and $S_2$ are two
noncomplementary parts (they need not be identical)
that together form the system $S$. Rest of $U$ is
the environment $E$. (b) Chain: the symmetrically placed (about dotted $C_2$
axis) filled circles represent two parts/sites under study.}
\label{univ}
\end{figure}

Let $S_1$ and $S_2$ be any two noncomplementary parts of a total system $U$.
The rest of the system is called the environment ($E$), which generally
consists of  many sites (Fig. \ref{univ}a). A measurement on $E$ by some
basis set would result in $S$ (= $S_1$ + $S_2$) assuming different pure
states with appropriate probabilities. Unlike in the case of {\it localizable
entanglement} \cite{popp} where only local measurements on the {\it individual}
sites of $E$ are allowed, we allow all possible measurements (including
the joint measurements on the sites) in our localization process.
It may be noted here that, all the measurements in this work are considered to
be non-selective projective-type. In the next part, we derive a simple but
generic optimality condition for the LE, which will
help us find optimal (more generally, stationary) solutions and check whether a
given solution is optimal. Studying a general bipartite system in mixed state
is notoriously difficult. There can be innumerable ways of decomposing a mixed
state, where each decomposition corresponds to an average entanglement
(entropy). The maximum and the minimum possible values of the average
entanglement are termed as {\it entanglement of assistance} (EoA)
\cite{vincenzo} and {\it entanglement of formation} (EoF) \cite{bennet}
respectively. The optimality condition derived here may be of use in finding
them. In this regard, a brief discussion is given after arriving at the
condition.

{\it The optimality condition.}- When expressed in the product basis states of
$E$ and $S$ (Fig.~\ref{univ}a), the given wave function
(that we study) becomes, $|\Psi\rangle = \sum_{i,j=1,1}^{D_E,D_S}C_{i,j}
|\xi_i\rangle^E |\phi_j\rangle^S$. Here $|\xi\rangle^E$s
($|\phi\rangle^S$s) are some orthonormal basis vectors of the state space of
$E$ ($S$) with dimensionality $D_E$ ($D_S$). The state can also be written as
\begin{eqnarray}
\label{wvfn}
|\Psi\rangle = \sum\nolimits_{i=1}^{D}
\sqrt{p_i}|\xi_i\rangle^E |\xi_i\rangle^S, 
\end{eqnarray}
with $p_i = \sum_{j'=1}^{D_S}C_{i,j'}C_{i,j'}^*$ and $|\xi_i\rangle^S =
\sum_{j=1}^{D_S}\frac{C_{i,j}}{\sqrt{p_i}}
|\phi_j\rangle^S$. Here the summation runs over nonzero $p_i$'s, numbering
$D$ ($\le D_E$). In general, states $|\xi\rangle^S$s are not orthonormal.
The operational interpretation of the later expression of the state
$|\Psi\rangle$ is that, if we perform measurement
on $E$ by the basis set $\{\xi^E\}$, the state will collapse
and we will get $S$ in different pure states $|\xi_i\rangle^S$s  with
corresponding probabilities $p_i$'s.

If $\mathcal{S}_i$ be the entropy of $|\xi_i\rangle^S$, then the
average entropy (entanglement) localized between $S_1$ and $S_1$ would be,
\begin{eqnarray}
\label{aventr}
\bar{\mathcal{S}}\{\xi^E\} = \sum\nolimits_{i=1}^{D} 
p_i \mathcal{S}_i.
\end{eqnarray}
As both $p_i$'s and $\mathcal{S}_i$'s depend on the choice of the BSM,
the average entropy (or LE) $\bar{\mathcal{S}}$ will also depend on the choice
of the BSM. We need to derive a condition for the choice of BSM
($\{\xi^E\}$) which optimizes $\bar{\mathcal{S}}$.

We first note that, any general basis set can be obtained from an initial basis
set $\{\xi^E\}$ by application of a series of elementary transformations (ETs).
Here an ET is a small-angle orthonormal transformation between any two initial
basis states keeping others unchanged. We now derive first
order change in $\bar{\mathcal{S}}$ due to an ET. If $|\xi_i\rangle^E$ and
$|\xi_j\rangle^E$ be any two initial basis states, then the two new basis states
obtained by an ET would be,
\begin{eqnarray}
\label{newbasis}
|\xi'_i \rangle^E =|\xi_i\rangle^E +\epsilon |\xi_j\rangle^E ~{\rm and}~
|\xi'_j \rangle^E =|\xi_j\rangle^E -\epsilon |\xi_i\rangle^E.
\end{eqnarray}
Here $\epsilon$ is the small angle (a parameter) whose higher order
terms can be neglected. Due to change in these basis states, corresponding
probabilities and states of the $S$ would also change (see eq.~\ref{wvfn}).
We need to relate these new probabilities and states with the old ones.

At this stage it is advantageous to express all the probabilities as the
diagonal elements of a density operator (matrix), which is, in our case,
the reduced density matrix (RDM) of $E$ ($\rho^E$). The elements
of the RDM are given by
$\rho^E_{ii'}=\sum_{j=1}^{D_S}C_{i,j}C_{i',j}^*$.
Using this RDM, probability corresponding to a state $|\xi\rangle^E$ would be
$p=~^E\langle \xi| \rho^E |\xi\rangle^E$. This allows us to write the new
probabilities as (using eq.~\ref{newbasis}),
\begin{eqnarray}
\label{newprbs}
p'_i=p_i+\epsilon k_{ij} ~{\rm and}~ p'_j=p_j-\epsilon k_{ji},
\end{eqnarray}
with $k_{ij}=k_{ji}=~^E\langle \xi_i|\rho^E|\xi_j\rangle^E+~^E\langle \xi_j
|\rho^E| \xi_i\rangle^E$. Let us first consider the case when none of the
$p_i$ and $p_j$ is zero.
Now if $|\xi'_i \rangle^S$ and $|\xi'_j \rangle^S$ be the new states of the
$S$, then, in the new scenario, the state $|\Psi\rangle$ can be rewritten as,
\begin{eqnarray}
\label{wvfn1}
|\Psi\rangle = \sqrt{p'_i}|\xi'_i\rangle^E |\xi'_i\rangle^S+\sqrt{p'_j}
|\xi'_j\rangle^E |\xi'_j\rangle^S+\cdots
\end{eqnarray}
Here we focus only on $i$-th and $j$-th states, as other states are unchanged.
Now using eqns. \ref{newbasis} and \ref{newprbs} in the above expression and
then comparing the terms associated with the initial basis states
$|\xi_i\rangle^E$ and $|\xi_j\rangle^E$ from the two different expressions of
$|\Psi\rangle$ (in eqs.~\ref{wvfn} and~\ref{wvfn1}), we get the following
solutions for the new states of $S$:
\begin{eqnarray}
\label{newst1}
|\xi'_i\rangle^S = |\xi_i\rangle^S+\epsilon\left(a_{ij}|\xi_i\rangle^S+b_{ij}
|\xi_j\rangle^S\right)\\
\label{newst2}
|\xi'_j\rangle^S = |\xi_j\rangle^S-\epsilon\left(a_{ji}|\xi_j\rangle^S+b_{ji}
|\xi_i\rangle^S\right)
\end{eqnarray}
Here $a_{ij}= -\frac{1}{2}k_{ij}p_i^{-1}$ and $b_{ij}= p_j^{1/2}p_i^{-1/2}$
in eq.~\ref{newst1}. Interchanging the indices $i$ and $j$ we get
similar terms in eq.~\ref{newst2}.

Now let $Q=\{Q_{lm}\}$ and $R=\{R_{lm}\}$ be the matrices representing
respectively the states $|\xi_i\rangle^S$ and $|\xi_j\rangle^S$ in some product
basis states of the parts $S_1$ and $S_2$. In terms of these matrices, the RDMs
for $S_1$ would be $\rho^{S_1}(\xi_i)=Q Q^{\dagger}$ and
$\rho^{S_1}(\xi_j)=R R^{\dagger}$ while $S$ is respectively in
$|\xi_i\rangle^S$ and $|\xi_j\rangle^S$.
Similarly, the RDMs for $S_1$ corresponding to the new states, given in
eqs.~\ref{newst1} and~\ref{newst2}, would be,
\begin{eqnarray}
\label{newrho1}
\rho^{S_1}(\xi'_i) = \rho^{S_1}(\xi_i)+\epsilon \left(2a_{ij}\rho^{S_1}(\xi_i)+
2b_{ij}\Delta_{ij}\right)\\
\label{newrho2}
\rho^{S_1}(\xi'_j) = \rho^{S_1}(\xi_j)-\epsilon \left(2a_{ji}\rho^{S_1}(\xi_j)+
2b_{ji}\Delta_{ji}\right)
\end{eqnarray}
Here $\Delta_{ij} = \frac{1}{2}(QR^{\dagger}+RQ^{\dagger})$, a Hermitian
matrix. Let us denote here the changes in the RDMs in
eqs.~\ref{newrho1} and~\ref{newrho2}  as  $\epsilon\rho^{S_1}_1(ij)$ and
$-\epsilon\rho^{S_1}_1(ji)$ respectively. It is worth mentioning that, as
trace (Tr) of any RDM is 1, we have
\begin{eqnarray}
\label{tr_rho1}
{\rm Tr}~ \rho^{S_1}_1(ij) = {\rm Tr}~\rho^{S_1}_1(ji)=0.
\end{eqnarray}

We now use the relation $\rho^{S_1}(\xi'_i)~{\rm log_2~} \rho^{S_1}(\xi'_i) = 
\rho^{S_1}(\xi_i)~{\rm log_2~} \rho^{S_1}(\xi_i) +\epsilon \rho^{S_1}_1(ij) +
\epsilon \rho^{S_1}_1(ij) ~{\rm log_2~} \rho^{S_1}(\xi_i)$ \cite{1stexp}
for obtaining entropy corresponding to the new state
$|\xi'_i\rangle^S$. This operator relation is not ill-defined due to last
term as both
$\rho^{S_1}(\xi_i)$ and $\rho^{S_1}_1(ij)$ go to zero simultaneously (this can
be understood by singular value decomposition of the matrix Q). Now tracing
over both sides of this relation and a similar relation for the
$j$-th state, we respectively get the following entropies for the
new states of $S$,
\begin{eqnarray}
\label{new_entr1}
\mathcal{S}'_i&=&\mathcal{S}_i-\epsilon{\rm Tr}~\rho^{S_1}_1(ij)~{\rm log_2~} 
\rho^{S_1}(\xi_i)~~{\rm and}~\\
\label{new_entr2}
\mathcal{S}'_j&=&\mathcal{S}_j+\epsilon{\rm Tr}~\rho^{S_1}_1(ji)~{\rm log_2~} 
\rho^{S_1}(\xi_j).
\end{eqnarray}
Here we used eq.~\ref{tr_rho1} to get these relations. Let us now denote the
changes in entropies in eqs.~\ref{new_entr1} and~\ref{new_entr2} as
$-\epsilon\mathcal{S}^1_{ij}$ and $\epsilon\mathcal{S}^1_{ji}$ respectively.

Now using these new entropies along with the new probabilities in eqn.
\ref{newprbs}, we get the new average entropy:
\begin{eqnarray}
\label{newaventr}
\bar{\mathcal{S}}'=\sum\nolimits_{l=1}^{D} p'_l \mathcal{S}'_l
=\bar{\mathcal{S}}+\epsilon \bar{\mathcal{S}}_1, 
\end{eqnarray}
where,
$\bar{\mathcal{S}}_1=k_{ij}\mathcal{S}_i-p_i\mathcal{S}^1_{ij}-
k_{ji}\mathcal{S}_j+p_j\mathcal{S}^1_{ji}$.

Before we set the optimality condition, we now check the cases when both are
or one of $p_i$ and $p_j$ is zero. When $p_i=p_j=0$, then $k_{ij}=k_{ji}=0$.
Therefore, $p'_i=p'_j=0$ (see eq. \ref{newprbs}). Which
implies that $\bar{\mathcal{S}}_1$ is zero. On the other hand, when $p_i\neq0$
and $p_j=0$, we gave again $k_{ij}=k_{ji}=0$. From eq. \ref{newprbs} we have
$p'_i = p_i$ and $p'_j=0$. Now it is clear from eq. \ref{wvfn1} (with second
term being zero) that, this type of ETs are not allowed (within the first
order calculation).

So, the desired {\it optimality condition} is
$\bar{\mathcal{S}}_1=0$ or,
\begin{eqnarray}
\label{cndtn1}
k_{ij}\mathcal{S}_i-p_i\mathcal{S}^1_{ij}=k_{ji}\mathcal{S}_j-
p_j\mathcal{S}^1_{ji},
\end{eqnarray}
for all $i$ and $j$ for which corresponding probabilities are
nonzero. The second order change in $\bar{\mathcal{S}}$ due to different ETs
can also be derived but they can not confirm actual character of an optimum
\cite{sahoo}.

For the derivation of the above optimality condition, we have assumed the
existence of a fixed environment ($E$). Therefore, the condition will be helpful
when we have a definite tripartite system and we want to localize entanglement
between two parts in an optimal way by performing measurement on the third
part. Now the question is whether it also can be of any use for calculating
EoA and EoF of a mixed state. We here note that, for a bipartite
system in a mixed state, it is always possible to construct
a pure state by augmenting the bipartite system with an
ancilla in such a way that the RDM of
the system becomes the given mixed state. By performing all possible
measurements on all possible ancillas, we get all possible
decompositions of the mixed state. By a theorem of Hughston-Jozsa-Wootters
\cite{hughston}, we can relate each decomposition to an $M_{r\times k}$ matrix
with $k$ orthonormal column vectors. Here the $r$ and the $k$ are the number of
terms in the decomposition and the rank of the mixed state respectively. This
implies that, for a particular ancilla, we can express the optimality condition
in terms of a matrix $M_{r\times k}$. Now as in principle number of terms in a
decomposition can be anything, it may appear that the optimality condition
obtained for a fixed ancilla would be of no use for the said purpose.
Fortunately, at least in the case of EoF, it is seen that,
consideration of a few number of terms in the decomposition is enough for
extremization \cite{wootters}. Hence we hope that, optimality condition given
here may also be useful in calculating EoA and EoF of a mixed state.

{\it The LECM.}-  Our canonical way of localizing entanglement between $S_1$
and $S_2$ is to take eigenstates of $\rho^E$ (the RDM of $E$) as the BSM and
perform measurement on $E$. We will find now the expression for the
entanglement localized in this way (which we call LECM).

Expression of the LECM can easily be obtained from Schmidt decomposition (SD)
of the state under study. The SD of the state $|\Psi\rangle$ into the product
states of $E$ and $S$ is given by,
\begin{eqnarray}
\label{sd1}
|\Psi\rangle = \sum\nolimits_i \sqrt{\lambda_i} |\underline{\xi}_i\rangle^S 
|\underline{\xi}_i\rangle^E.
\end{eqnarray}
Here, $|\underline{\xi}_i\rangle^S$ ($|\underline{\xi}_i\rangle^E$) is the
$i$-th eigenstate of the RDM of $S$ ($E$) corresponding to the eigenvalue
$\lambda_i$. The index $i$ runs from 1 to $D_{sn}$ (Schmidt number). An
operational interpretation of eq. \ref{sd1} is as follows: measurement on
$E$ by the basis set $\{\underline{\xi}^E\}$ would result in $S$ assuming
state $|\underline{\xi}_i\rangle^S$ with probability $\lambda_i$. Now if
$\underline{\mathcal{S}}_i$ is the entropy of $|\underline{\xi}_i\rangle^S$
then, we can write directly from eq. \ref{aventr},
\begin{eqnarray}
\label{entr_av}
\bar{\mathcal{S}}=\sum\nolimits_{i=1}^{D_{sn}} \lambda_i 
\underline{\mathcal{S}}_i.
\end{eqnarray}
This is the desired expression for the LECM.

This localization procedure may appear to face some problems when $\rho^S$ has
degenerate eigenstates. Most of the time this difficulty can easily be resolved
by the use of conserved quantities and the symmetries of the system
\cite{sahoo}.

We now discuss the important cases when the LECM can be shown to be optimal. We
note that when measurement is performed by eigenstates of RDM
($\rho^{E}$), $k_{ij}=k_{ji}=0$ and the  eq.~\ref{cndtn1} reduces to
\begin{eqnarray}
\label{1cndtn_rdm}
 {\rm Tr}~\Delta_{ij}~
{\rm log_2~} \rho^{S_1}(\underline{\xi}_i)={\rm Tr}~\Delta_{ji}~
{\rm log_2~} \rho^{S_1}(\underline{\xi}_j). 
\end{eqnarray}

For spin-1/2 systems, with both the parts $S_1$ and $S_2$ taken to be
single-sites,
use of conserved quantity (here $Z$-component of total spin) and parity or $C_2$
symmetry (for finite systems; for translationally invariant systems this
will be automatically satisfied) leads the four eigenstates of $\rho^{S}$ into the
following form: $|\uparrow\uparrow\rangle$,
$\frac{1}{\sqrt{2}}(|\uparrow\downarrow\rangle\pm|\downarrow\uparrow\rangle)$
and $|\downarrow\downarrow\rangle$, with $\uparrow$ ($\downarrow$) being up
(down) spin.  Any two of the $2\times 2$ matrices representing these
states are seen to satisfy the condition given in eq. \ref{1cndtn_rdm}.
Since entropies of first and fourth states are zero, and second and third
states (a singlet and a triplet) are one, the LECM in this case simply
becomes,  $\bar{\mathcal{S}}=\lambda_s+\lambda_t$, with $\lambda_s$
($\lambda_t$) is the eigenvalue of $\rho^{S}$ corresponding to singlet
(triplet). Note, after measurement, two sites will be in the statistical
mixture of all the four eigenstates of $\rho^{S}$ and if $\lambda_s$ =
$\lambda_t$, then we will not be able to extract any useful entangled pair
from the ensemble.

{\it Numerical Result.}- We study the LECM ($\bar{\mathcal{S}}$) between two
sites of a frustrated antiferromagnetic
Heisenberg chain ($j_1-j_2$ spin-1/2 model \cite{majumdar}). Two sites
are placed symmetrically as in fig.~\ref{univ}b; this arrangement makes
$\bar{\mathcal{S}}$ to be an optimal.

\begin{figure}[t]
\includegraphics[width=8.0cm]{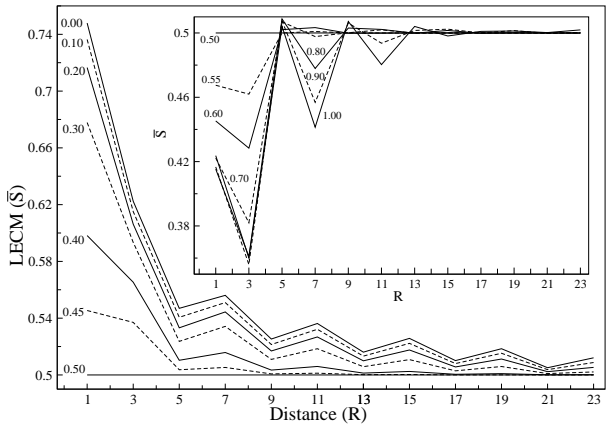}
\caption{\small For the groundstate of a $j_1-j_2$ spin chain with
24 sites, the LECM ($\bar{\mathcal{S}}$) against distance between two sites
($R$) is shown for different $j_2$ values. The special case $j_2 = 0$ is just
result for an antiferromagnetic spin-1/2 Heisenberg chain. In the inset, result
is shown for $j_2\ge0.5$ (the spiral phase).}
\label{qpt_j1j2}
\end{figure}

Its behavior
against the distance between the sites ($R$) for different values of $j_2$ is
shown in fig.~\ref{qpt_j1j2}.

\begin{figure}[t]
\includegraphics[width=8.0cm]{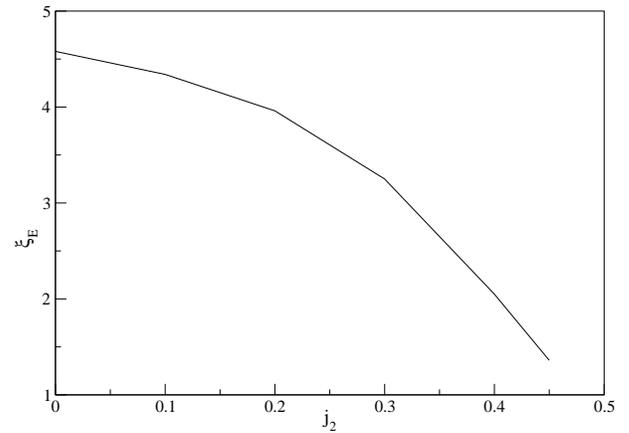}
\caption{\small The {\it entanglement length} $\xi_E$ is shown here as a
function of $j_2$.}
\label{enlng}
\end{figure}

We see that, in the  N\'{e}el phase ($j_2<0.5$), $\bar{\mathcal{S}}$
falls with increasing $R$ and reaches a constant value at large $R$.
With increase in $R$, all the four eigenvalues of the $\rho^{S}$ become
equal (hence $\lambda_s$ = $\lambda_t$), which results in the LECM assuming a
value of 0.5 (which we call residual value or $\bar{\mathcal{S}}_r$).
In fact, this particular value of the LECM is obtained
if we take two sites (symmetrically) one each from two totally separate
chains (unentangled) and perform canonical measurement on remaining parts
of the chains. Physically this implies that, in case of a single chain (where
sites are connected), when $R$ is large, two sites
become unentangled, i.e., we can not localize `extractable' or useful
entanglement between them by a canonical measurement.
Keeping this in mind, we therefore, can quantify actual extractable entanglement
in our localization process as $\Delta\bar{\mathcal{S}} = 
\bar{\mathcal{S}}-\bar{\mathcal{S}}_r$. Any positive value of the quantity
$\Delta\bar{\mathcal{S}}$ will give us the actual `gain' in our localization
process. In case of our single chain, the
quantity $\Delta\bar{\mathcal{S}}$ falls with increasing $R$, which can
be understood qualitatively by the Valence Bond (VB) theory \cite{soos,sahoo1}.
A groundstate can be expressed by linear combination of many VB basis states
where basis states with nearest neighbor bonds (a bond represents an entangled
pair) contribute more towards the groundstate compared to the ones with distant
neighbor bonds. This says why for large $R$ two sites become decoherent or
unentangled. This fact naturally leads us to the notion of {\it entanglement
length} ($\xi_E$), which is the typical length scale upto which it is possible
to localize useful or extractable entanglement between two sites.
If fall in $\Delta\bar{\mathcal{S}}$ with increasing $R$ is assumed to be
exponential in nature, we can define $\xi_E$ as \cite{popp,aharonov},
$\xi^{-1}_E \rightarrow -\frac{{\rm ln}\Delta\bar{\mathcal{S}}}{R}$ for
large $R$. Since we have a finite system
(24 sites), we can use two values of $R$ and corresponding values of
$\Delta\bar{\mathcal{S}}$ to estimate the value of $\xi_E$ in the following
way: $\xi^{-1}_E = - \frac{{\rm ln}\Delta\bar{\mathcal{S}}_1 - 
{\rm ln}\Delta\bar{\mathcal{S}}_2} {R_1 -R_2}$.
We take $R$ = 7 and 11 for this purpose (this particular values are
chosen to avoid odd-even effect of a finite chain \cite{sahoo1}), and
calculated values of $\xi_E$ as a function of $j_2$ can be seen from
fig.~\ref{enlng}.

The fall in the value of $\xi_E$ with increasing value of $j_2$ is not
unexpected, as the contribution of VB basis states with long bonds
decreases with the increasing value of $j_2$. This is supported by the fact
that at the Majumdar-Ghosh (MG) point ($j_2 = 0.5$) groundstate has only
nearest neighbor bonding.

The large oscillations in the value of $\bar{\mathcal{S}}$ for $j_2>0.5$
(fig.~\ref{qpt_j1j2}) can be understood by the fact that the phase of the
system in this range is spiral in nature. The degree of entanglement between
two sites depends on the relative phase factor between the sites. This is why
for some distances the value of $\bar{\mathcal{S}}$ is very low.

{\it Conclusion.}-In this letter, we have derived a simple but generic
optimality condition, which would be helpful in finding optimal values of the
entanglement localized between two disjoint parts of a many-body system by doing
measurement on the remaining part of the system. Besides, we also discussed how
it can be useful in studying mixed states of a general bipartite system. We
further discussed a canonical way of localizing entanglement which in some
important and not-too-restricted cases (shown for spin-1/2 systems)
gives optimal value of the
localized entanglement. The entanglement localized by canonical measurement
(or LECM) is operationally meaningful and easy to calculate in practice.
Unlike other measurement-based quantifications of entanglement, the
LECM does not require any demanding optimization procedure. Another important
advantage of this LECM is that, since it does not depend on arbitrary choice of BSM,
it provides a general framework for comparative study of different types of quantum
many-body systems. We studied
a $j_1-j_2$ spin model to demonstrate the behavior of LECM. In this context, we
also discussed extractable or useful part of LECM and defined an entanglement
length scale upto which one can localize extractable entanglement between two
sites. It may be stressed here that, all the
concepts in this letter are quite general, virtually applicable to any
kind of quantum many-body systems.

I thank Prof. S. Ramasesha (SR) and Prof. Diptiman Sen 
for useful discussions. I also acknowledge SR's financial support through his 
project from DST, India.


\begin{thebibliography}{}
\bibitem{1stexp}If $\lambda_l$ be the $l$-th eigenvalue of $\rho^{S_1}(\xi_i)$
and $\lambda^1_l$ be the expectation value of $\rho^{S_1}_1(ij)$ in the
eigenstate, then the value of the operator $\rho^{S_1}(\xi'_i)~
{\rm log_2~} \rho^{S_1}(\xi'_i)$ in the state would be
$(\lambda_l+\epsilon \lambda^1_l)~{\rm log_2~}\lambda_l(1+\epsilon 
\frac{\lambda^1_l}{\lambda_l})$. This equals to $\lambda_l~{\rm log_2~}
\lambda_l + \epsilon \lambda^1_l + \epsilon \lambda^1_l ~{\rm log_2~}
\lambda_l$, using ${\rm log_2~}(1+\epsilon \frac{\lambda^1_l}
{\lambda_l}) = \epsilon \frac{\lambda^1_l}{\lambda_l}$. This eventually
suggests the operator relation we use.
\end{thebibliography}
\end{document}